\documentclass[prl,aps,twocolumn]{revtex4}
\usepackage{epsfig,amsmath}
\usepackage{bm}
\def\Fbox#1{\vskip1ex\hbox to 8.5cm{\hfil\fboxsep0.3cm\fbox{%
\parbox{8.0cm}{#1}}\hfil}\vskip1ex\noindent}

\newcommand{\C}[1]{{\mathcal{#1}}}

\newcommand{\Onecol} {\begin{widetext} \onecolumngrid} 
\newcommand{\Twocol} {\end{widetext} \twocolumngrid} 

\newcommand{\be}{\begin{equation}}
\newcommand{\ba}{\begin{array}}
\newcommand{\bea}{\begin{eqnarray}}
\newcommand{\bfi}{\begin{figure}}
\newcommand{\ee}{\end{equation}}
\newcommand{\ea}{\end{array}}
\newcommand{\eea}{\end{eqnarray}}
\newcommand{\efi}{\end{figure}}

\def\DR{{\rm D\!R}}
\def\pDR{{\%\!\DR}}
\def\Re{${\C R}\mkern-3.1mu e$} 
\def\RE{{\C R}\mkern-3.1mu e} 
\renewcommand{\sb}[1]{_{\text {#1}}} 
\newcommand{\Sb}[1]{_{_{\text {#1}}}} 
\begin{document}
\title{Saturation of Turbulent Drag Reduction in Dilute Polymer Solutions}
\author{Roberto Benzi$^\dag$, Victor S.
L'vov$^*$, Itamar Procaccia$^*$ and Vasil Tiberkevich$^*$}
\affiliation{$^\dag$ Dip. di Fisica and INFM, Universit\`a ``Tor
Vergata", Via della Ricerca Scientifica 1, I-00133 Roma, Italy,\\
$^*$Dept. of Chemical Physics, The Weizmann Institute of Science,
Rehovot, 76100 Israel} 
\begin{abstract}
Drag reduction by polymers in turbulent wall-bounded flows exhibits universal and non-universal
aspects. The universal maximal mean velocity profile was explained in a recent theory. The saturation of this
profile and the crossover back to the Newtonian plug are non-universal, depending on Reynolds number \Re, concentration
of polymer $c_p$ and the degree of polymerization $N_p$. We explain the mechanism of saturation stemming
from the finiteness of extensibility of the polymers,
predict its dependence on $c_p$ and $N$ in the limit of small $c_p$ and large Re, and present the
excellent comparison of our predictions to experiments on drag reduction by DNA.

\end{abstract}
\maketitle 

\begin{figure}[t]\centering
\includegraphics[width=0.5\textwidth]{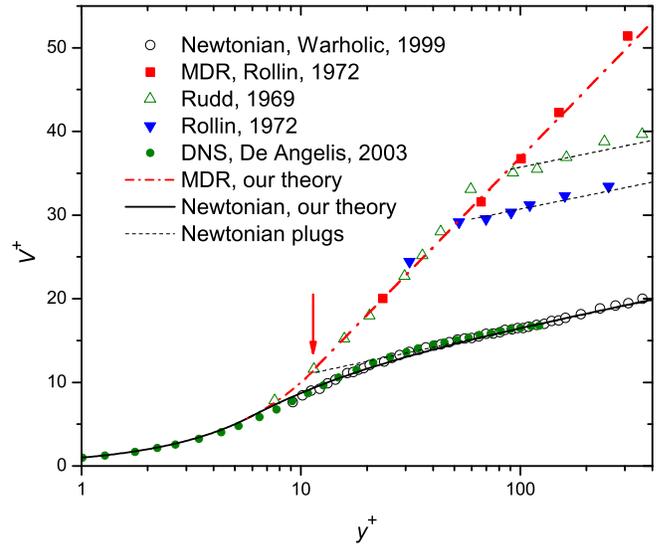}
\caption{Mean normalized  velocity profiles as a function of the
normalized  distance from the wall during drag reduction. The data
points from numerical simulations (green circles) \cite{03ACLPP}
and the experimental points (open circles) \cite{99WMH} represent
the Newtonian results.  The red data points (squares) \cite{72RS}
represent the Maximum Drag Reduction (MDR) asymptote. The dashed
red curve represents the theory of \cite{03LPPT} which agrees with
the universal law (\ref{log-p}). The arrow marks the crossover
from the viscous linear law (\ref{viscous}) to the  asymptotic
logarithmic law (\ref{log-p}). The blue filled triangles
\cite{72RS} and green open triangles \cite{69Rud} represent the
crossover, for intermediate concentrations of the polymer,  from
the MDR asymptote to the Newtonian plug. This Letter presents a
theory of this crossover.} \label{profiles}
\end{figure}

\noindent{\bf Introduction}. The onset of turbulence in fluid
flows is accompanied by a significant increase in the drag. This
drag poses a real technological hindrance to the transport of
fluids and to the navigation of ships. It is interesting therefore
that the addition of long chained polymers to wall-bounded
turbulent flows can result in a significant reduction in the drag
\cite{49Tom}. The basic experimental knowledge of the phenomenon
had been reviewed and systematized by Virk \cite{75Vir}; the
increase in mean velocity profile as a function of the distance
from the wall depends on the characteristics of the polymer and
its concentration, but cannot exceed a universal asymptote known
as the ``Maximum Drag Reduction" (MDR) curve which is independent
of the polymer's concentration or its characteristics. When the
concentration is not large enough, the mean velocity profile
follows the MDR for a while and then crosses back to a
Newtonian-like profile, known as the ``Newtonian plug" cf. Fig.
\ref{profiles}. Recently the nature of the MDR and the mechanism
leading to its establishment were explained \cite{03LPPT}. For
Newtonian flows  the momentum flux is dominated by the so-called
Reynolds stress, leading to a logarithmic (von-Karman) dependence
of the mean velocity on the distance from the wall \cite{00Pope}.
With the addition of polymers, while momentum is produced at a
fixed rate by the forcing, polymer stretching results in a
suppression of the Reynolds stress (and thus of the momentum flux
from the bulk to the wall). Accordingly the mean velocity in the
channel must increase. It was shown that when the concentration of
the polymers is large enough there exists a new logarithmic law
for the mean velocity with a slope that fits existing numerical
and experimental data. The law is universal, thus explaining the
MDR asymptote. It turned out that the polymer stretching leads to
an effective viscosity which increases with the distance from the
wall. This effective viscosity attains a {\em self-consistent
profile}, increasing {\em linearly} with the distance from the
wall.  With this profile the reduction in the momentum flux from
the bulk to the wall overwhelms the increased dissipation that
results from the increased viscosity. Thus the mean momentum
increases in the bulk, and this is how drag reduction is realized.
In \cite{04DCLPT} it was demonstrated by DNS that Navier-Stokes
flows with viscosity profiles that vary linearly with the distance
from the wall indeed show drag reduction in close correspondence
with the phenomena seen in full viscoelastic simulations.

\begin{figure}
\centering
\includegraphics[width=0.45\textwidth]{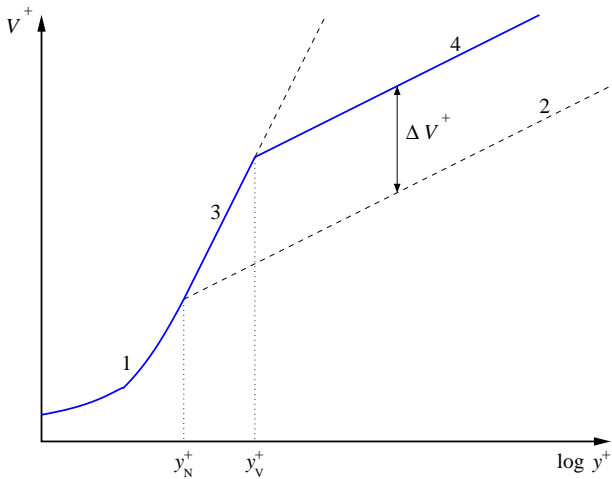}
\caption{Schematic mean velocity profiles. Region 1: $y^+ <
y^+\Sb N$, -- viscous sublayer.  Region 2: $ y^+ > y^+\Sb N $, --
logarithmic layer for the turbulent Newtonian flow. Region 3: $
y^+\Sb N  < y^+ < y^+ \Sb V =  (1+Q) \, y^+\Sb N$ -- MDR
asymptotic profile in the viscoelastic flow.  Region 4: $ y^+
> y^+ \Sb V $ -- Newtonian plug in the viscoelastic flow.
 \label{schemprof}}
\end{figure}

\noindent {\bf Theory of the saturation of drag reduction. }
The aim of this Letter is to provide a theory of the crossovers
from the MDR to the Newtonian plug  for small polymer
concentration $c\sb p$ and large \Re. We show that  the mechanism
for crossover in that limit is the saturation of the increase of
effective viscosity: at some point there is no enough polymer to
supply the necessary linear profile of effective viscosity. We
will develop the theory for a channel geometry, and then apply the
results to a rotating disk experiment with DNA as the drag
reducing agent. For a channel of width $2L$ we denote by $x$, $y$
and $z$ the streamwise, the wall normal and the spanwise
directions respectively. The only non-zero mean velocity component
is $V(y)$ in the streamwise direction. The traditional wall units
are
\begin{equation}
\RE_\tau \equiv {L\sqrt{\mathstrut p' L}}/{\nu_0}\ , \ y^+ \equiv
{y \RE_\tau }/{L} \ , \ V^+ \equiv {V}/{\sqrt{\mathstrut p'L}} \ .
\label{red}
\end{equation}
Here $p'\equiv -\partial p/\partial x$ is the  pressure gradient
in the streamwise direction $x$. $\RE_\tau$ is known as the
friction Reynolds number. $\nu_0$ is the kinematic viscosity of
the neat fluid. In terms of these variables, the predictions of
the standard Newtonian theory and the theory of \cite{03LPPT} are
summarized with the help of Fig. \ref{schemprof} in which
the Newtonian and viscoelastic velocity profiles are shown
schematically in the reduced coordinates~(\ref{red}). In the
Newtonian flow there are two distinct regions: a viscous sublayer
(region 1) with the reduced velocity given by
\begin{equation}\label{viscous}
  V^+(y^+) = y^+ \,,\end{equation} and a logarithmic layer (region
2), in which velocity is given by Prandtl-Karman law:
\begin{equation}
\label{log-n}  V^+(y^+) = \kappa\Sb N^{-1}\log y^+ +B\Sb N ,~
\text{(Prandtl-Karman law).}
\end{equation}
Here $\kappa\Sb N^{-1}\simeq2.29$ and $B\Sb N\simeq6.13$
\cite{97ZS}.
The mean velocity profile in the viscoelastic case  consists of
three regions \cite{75Vir}: a viscous sublayer, a logarithmic
polymeric sublayer (region 3 in the Fig.~\ref{schemprof}) with the
slope greater then the Newtonian one:
\begin{equation}\label{log-p}
  V^+(y^+) = \kappa\Sb V^{-1}\log y^+ +B\Sb V \ , \text{(MDR profile),}
\end{equation}
($\kappa\Sb V^{-1}\simeq11.7$, $B\Sb V\simeq-17$ \cite{75Vir}),
and a Newtonian plug (region 4). In the last region the velocity
follows a log law with the Newtonian slope, but with some
 velocity increment $\Delta V^+$:
\begin{equation}\label{log-np}
  V^+ = \kappa\Sb N^{-1}\log y^+ +B\Sb N+\Delta V^+
\ .\end{equation} Note that the three profiles Eqs.
(\ref{viscous}), (\ref{log-n}), and (\ref{log-p}) intercept at one
point $y^+ = y^+\Sb N\simeq\kappa\Sb V^{-1}\simeq11.7$.

The increment $\Delta V^+$ which determines the amount of drag
reduction is in turn determined by the crossover from the MDR to
the Newtonian plug (see Fig.~\ref{schemprof}). We refer to this
cross over point as $y^+\Sb V$. To measure the quality of drag
reduction we introduce a dimensionless drag reduction parameter
\begin{equation}\label{Q}
 Q \equiv \frac{y^+\Sb V}{y^+\Sb N} -1\ .
\end{equation}
The velocity increment $\Delta V^+$ is related to this
parameter as follows
\begin{equation}\label{DeltaV}
  \Delta V^+ = \left(\kappa\Sb V^{-1}-\kappa\Sb N^{-1}\right)
    \log\left(y^+\Sb V/y^+\Sb N\right)
  = \alpha\log(1+Q)
\ .\end{equation}
 Here $\alpha\equiv\kappa\Sb V^{-1}-\kappa\Sb
N^{-1}\simeq9.4$. The Newtonian flow is then a limiting case of
the viscoelastic flow corresponding to $Q=0$.

The crossover point $y^+\Sb V$ is non-universal, depending on \Re,
the number of polymers per unit volume $c\sb p$, the chemical
nature of the polymer, etc. According to the theory of drag
reduction \cite{03LPPT}, the total viscosity of the fluid
$\nu_{\rm tot}(y^+)=\nu_0+\nu_p(y^+)$ [where $\nu_p(y^+)$
is the polymeric contribution to the viscosity] is linear in $y^+$
in the MDR region:
\begin{equation}\label{nu-tot}
  \nu_{\rm tot}(y^+) = \nu_0y^+/y^+\Sb N\,, \qquad
y^+\Sb N  < y^+ < y^+ \Sb V\ .
\end{equation}
When the concentration of polymers is small and \Re~ is large
enough, {\em the crossover to the Newtonian plug at} $y^+\Sb V$
{\em occurs when the polymer stretching can no longer provide the
necessary increase of the total fluid viscosity}. In other words,
in that limit  the crossover is due to the finite extensibility
of the polymer molecules. Obviously, the polymeric viscosity can
not be greater than $\nu_{p\max}$ which is the viscosity of the
fully stretched polymers. Thus the total viscosity is limited by
$\nu_0+\nu_{p\max}$. Equating $\nu_0+\nu_{p\max}$ and $\nu_{\rm
tot}(y^+\Sb V)$ gives us the crossover position
\begin{equation}
  y^+\Sb V = y^+\Sb N(\nu_0+\nu_{p\max})/\nu_0
\ .\end{equation}
It follows from Eq. (\ref{Q}) that the drag reduction parameter is determined
very simply by
\begin{equation}\label{sigma}
  Q = \nu_{p\max}/\nu_0 \ ,
  \quad\text{$c\sb p$ small, \Re~ large} \ .
\end{equation}

At this point we need to relate the maximum polymeric viscosity
$\nu_{p\max}$ to the polymer properties. To this aim we estimate
the energy dissipation due to a single, fully stretched, polymer
molecule. In a reference frame co-moving with the polymer's center
of mass the fluid velocity can be estimated as $ u\simeq r \nabla
u$ (the polymer's center of the mass moves with the fluid velocity
due to negligible inertia of the molecule). The friction force
exerted on the $i$-th monomer is estimated using Stokes law,
\begin{equation}
F_i\simeq\rho_0\nu_0 \, a\,\delta u_i=\rho_0\nu_0 \,a\, r_i \nabla
u\, ,
\end{equation}
 where $a$ is an effective hydrodynamic radius of one monomer
(depending on the chemical composition), and $r_i$ is the distance
of the  $i$-th monomer from the center of the mass. In a fully
stretched state $r_i\simeq a\,i$ (the monomers are aligned along a
line). The energy dissipation rate (per unit volume) is equal to
the work performed by the external flow
\begin{eqnarray}
  -\frac{dE}{dt} &\simeq& c\sb p\sum_{i=1}^{N\sb p}
   F_i\delta u_i  \simeq
    \rho_0\nu_0a^3c\sb p N\sb p^3(\nabla u)^2 \nonumber\\
    &\equiv& \rho_0\nu_{p\max}(\nabla u)^2 \ .
\end{eqnarray}
We thus can estimate
$\nu_{p\max}$:
\begin{equation}
  \nu_{p\max} = \nu_0 a^3 c\sb p N\sb p ^3 \ . \label{result}
\end{equation}
Finally, the drag reduction parameter $Q$ is given by
\begin{equation}\label{sigma1}
  Q = a^3c\sb p N\sb p ^3 \qquad\text{$c\sb p$ small, \Re~ large}
\ .\end{equation} 
This is the central theoretical results of this Letter, relating
the concentration $c\sb p$ and degree of polymerization $N\sb p $
to the increment in mean velocity $\Delta V^+$ via Eq.
(\ref{DeltaV}).

\begin{figure}
\centering
\includegraphics[width=0.45\textwidth]{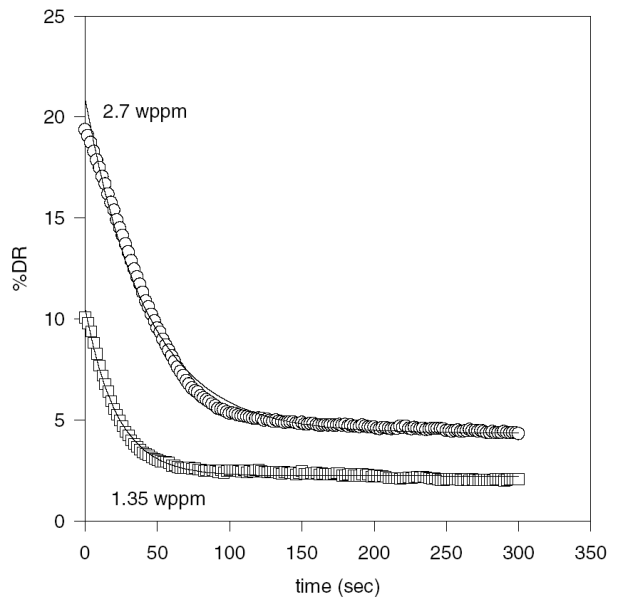}
\caption{$\pDR$ in a rotating disk experiment with $\lambda$-DNA
as the drag reducing polymer. Note that the $\pDR$ is proportional
to $c\sb p$. When the length $N\sb p$ reduces by a factor of 2
and, simultaneously, $c\sb p$ increases by factor of
$2$, the $\pDR$ reduces by a factor of 4.\label{Chan}}
\end{figure}

{\bf Testing and explaining a DNA experiment} \cite{02CLLC}.  A
particularly interesting experiment suitable for testing our
prediction was described in \cite{02CLLC}. Here turbulence was
produced in a rotating disk apparatus, with $\lambda$-DNA
molecules used to reduce the drag. The Reynolds number was
relatively high (the results below pertain to \Re~$\approx
1.2\times 10^6$) and the initial concentrations of DNA relatively
low (results employed below pertain to 2.70 and 1.35 wppm). During
the experiment DNA degrades; fortunately the degradation is very
predictable: double stranded molecules with 48 502 bp in size
degrade to double stranded molecules with 23 100 bp. Thus
invariably the length $N\sb p $ reduces by a factor of
approximately 2, and the concentration $c\sb p$ increases by a
factor of 2. The experiment followed the drag reduction efficacy
measured in terms of the percentage drag reduction defined by
\begin{equation}\label{def-DR}
  \pDR = \frac{T\Sb N -T\Sb V}{T\Sb N}\times 100
\,,\end{equation}
 where $T\Sb N$ and $T\Sb V$ are the torques
needed to maintain the disk to rotate at a particular Reynolds
number $\RE$ without and with polymers, respectively. The main
experimental results which are of interest to us are summarized in
Fig. \ref{Chan}.
We see from the experiment that both initially (with un-dergraded
DNA) and finally (with degraded DNA) the $\pDR$ is proportional to
$c\sb p$. Upon degrading, which amounts to decreasing the length
$N\sb p$ by a factor of approximately 2 and,
simultaneously increasing $c\sb p$  by factor of $2$,
 $\pDR$ decreases by a factor 4.

\begin{figure}
\centering\vskip.5cm
\includegraphics[width=0.45\textwidth]{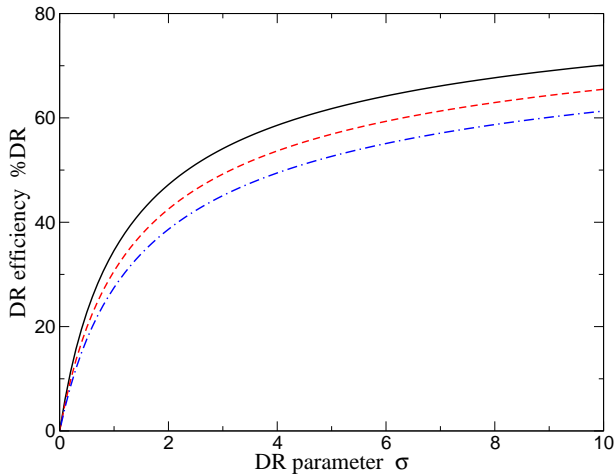}
\caption{Drag reduction efficiency $\pDR$ as function of the drag reducing
  parameter $Q$ for different Reynolds numbers $\RE$: $1.2\times10^5$,
  $1.2\times10^6$, and $1.2\times10^7$ (from top to bottom).\label{f:DR}}
\end{figure}

The flow geometry is rather complicated: with a rotating disk the
{\em linear} velocity depends on the radius, and the local \Re~is
a function of the radius. The drag reduction occurs however in a
relatively  small near-wall region, where the flow can be
considered as a flow near the flat plate. Thus, we consider an
equivalent channel flow -- with the same \Re~
 and a  half width $L$ of the order of height/radius of the
cylinder. In this plane geometry the torques in (\ref{def-DR})
should be replaced by the pressure gradients $p'\Sb{N,V}$:
\begin{equation}\label{def-DR1}
  \pDR = \frac{p'\Sb N -p'\Sb V}{p'\Sb N}\times 100 \ .
\end{equation}
In order to relate $\pDR$ with the drag reduction parameter $Q$,
we re-write Eq. (\ref{log-np}) in natural units
\begin{equation}
  V(y) =
    \sqrt{p'L}\left[
      \kappa\Sb N^{-1}\log\left(y\sqrt{p'L}/\nu_0\right)+B\Sb N+\Delta V^+
    \right]
\ .\end{equation} With constant \Re~the centerline velocity
$V_0=V(L)$ is kept fixed:
\begin{equation}\label{RE}
  \RE \equiv \frac{V_0L}{\nu_0} =
    \RE_\tau\left[\kappa\Sb N^{-1}\log\RE_\tau +B\Sb N +\Delta V^+\right]
\ .\end{equation}
 This equation implicitly determines the pressure gradient and
therefore the $\pDR$ as a function of $Q$ and ~\Re. The set of
Eqs.~(\ref{DeltaV}) and (\ref{RE}) is readily solved numerically,
and the solution for three different values of \Re~is shown in
Fig.~\ref{f:DR}. The middle curve corresponds to
$\RE=1.2\times10^6$, which coincides with the experimental
conditions \cite{02CLLC}. One sees, however, that the dependence
of $\pDR$ on the \Re~is rather weak.

One important consequence of the solutions shown in
Fig.~\ref{f:DR} is that for small $Q$ (actually for $Q\le 0.5$ or
$\pDR\le20$),  $\pDR$ is approximately a linear function of $Q$.
The experiments \cite{02CLLC} lie entirely within this linear
regime, in which we can linearize Eq.~(\ref{RE}) and find an
approximate solution for the $\pDR$:
\begin{equation}\label{DR-lin}
  \pDR =
    \frac{2\alpha Q}{\kappa\Sb N^{-1}\log(e\RE_{\tau}^0) +B\Sb N}\times100
\ .\end{equation}
Here $\RE_{\tau}^0$ is the friction Reynolds number for the Newtonian flow,
i.e. the solution of Eq.~(\ref{RE}) for $\Delta V^+=0$.

It is interesting to note, that while the $\pDR$ depends on the
Reynolds number, the ratio of different $\pDR$'s does not [to $O(
Q)$]:
\begin{equation}\label{DRrel}
  \frac{\pDR^{(1)}}{\pDR^{(2)}} =
  \frac{Q^{(1)}}{Q^{(2)}} =
  \frac{\nu_{p\max}^{(1)}}{\nu_{p\max}^{(2)}} \ .\end{equation}
This result, together with Eq. (\ref{result}), rationalizes
completely the experimental finding of \cite{02CLLC} summarized in Fig. \ref{Chan}.
During the DNA degradation, the concentration of polymers
increases by a factor of 2, while the number of monomers $N\sb p $
decreases by the same factor. This means that $\pDR$ should
decrease by a factor of 4, as is indeed the case.

{\bf Summary.} We have presented a theory of the crossover from
the universal MDR mean velocity profiles to the Newtonian plug. We
have connected the theory to measured percentages of drag
reduction, and tested the theory against experiments in which
$\lambda$-DNA is used as the drag-reducing agent. The experimental
results pertain to high \Re~and small $c\sb p$, where we can
assert that \emph{the crossover results from exhausting the
stretching of the polymers such that the maximal available
viscosity is achieved}. In the linear regime that pertains
to this experiment the degradation has a maximal effect on the
quality of drag reduction $Q$, leading to the precise factor of 4 in
the results shown in Fig. \ref{Chan}. Larger values of the
concentration of DNA will exceed the linear regime as is predicted
by Fig. \ref{f:DR}; then the degradation is expected to have a
smaller influence on the drag reduction efficacy. It is worthwhile
to test the predictions of this theory also in the nonlinear
regime. \vskip 0.5cm

\acknowledgments

We thank C.K. Chan for bringing Ref. \cite{02CLLC} to our attention and for
discussing the experimental results with us.
This work was supported in part by the US-Israel BSF, The ISF
administered by the Israeli Academy of Science, the European
Commission under a TMR grant and the Minerva Foundation, Munich,
Germany.

\end{document}